\documentclass[prb,aps,twocolumn,showpacs]{revtex4}

\usepackage{epsfig}

\begin{document}

\title{ $^{77}$Se NMR Evidence of Strongly Coupled Superconductivity in K$_{0.8}$Fe$_{2-x}$Se$_2$ }

\author{L. Ma ${^1}$}
\author{J. B. He ${^1}$}
\author{D. M. Wang ${^1}$}
\author{G. F. Chen $^{1} $}
\author{Weiqiang Yu $^{1}$}
\email{wqyu_phy@ruc.edu.cn}

\affiliation{
$^1$Department of Physics, Renmin University of China, Beijing 100872, China\\
}
\date{\today}

\begin{abstract}

We report the $^{77}$Se NMR Knight shift and spin-lattice relaxation studies on the superconducting state of the ternary iron selenide  
K$_{0.8}$Fe$_{2-x}$Se$_2$ with $T_c\approx$ 30 K.  Just below $T_c$, the Knight shift $^{77}K_n$ shows an immediate drop, indicating a singlet 
pairing. The spin-lattice relaxation rate $1/^{77}T_1$ decreases rapidly in the temperature range from $T_C$ to $T_c/2$, which can be fit with an 
isotropic gap of $\Delta \approx 3.8\pm 0.5 k_BT_c$. The Hebel-Slichter coherence peak is not observed. These data give bulk evidence for a strongly 
coupled superconductivity with isotropic gaps in  K$_{0.8}$Fe$_{2-x}$Se$_2$, which is similar to other iron-based high temperature superconductors. 
Below $T_c/2$, the spin-part of the Knight shift levels off to a constant value ($\sim 0.09\%$), and the spin-lattice relaxation follows a $1/T_1\sim 
T^{2}$ behavior, which are not well understood. 
  
\end{abstract}

\maketitle

The discovery of high-temperature superconductivity in iron pnictides \cite{Hosono_Jacs_130_3296, Chen_Nature_453_761, Chen_PRL_100_247002, 
Ren_MRI_12_105} has attracted a lot of research interests. Recently, high-temperature superconductivity is also observed in the ternary iron selenide 
A$_{y}$Fe$_{2-x}$Se$_2$ (A=K, Rb, Cs, {\it etc.}) \cite{Guo_PRB_82_180520, Chen_CM_10125525, FangMH_CM_10125236, Chen_CM_11010789} with $T_c\sim$ 32 
K, which is much higher than that of the binary iron selenide $\alpha$-Fe$_{1-\delta}$Se at the ambient pressure. Compared with other iron-based 
superconductors, K$_{0.8}$Fe$_{2}$Se$_2$ shows distinctive properties. First, iron vacancy seems to strongly affect the normal state metallic behavior 
and the superconductivity \cite{FangMH_CM_10125236, Chen_CM_11010789, WangNL_CM_11010572, LiJQ_CM_11012059, Zhang_CM_11012168}. Second, the system is 
probably heavily electron doped while the hole band is absent \cite{Feng_CM_10125980, Ding_EPL_83_47001, Yu_CM_11011017}. Theoretically, a band 
insulator \cite{Shein_CM_10125164}, and possibly with magnetic correlations \cite{Xiang_CM_10125536, Xiang_CM_10126015} has been suggested in 
stoichiometric AFe$_2$Se$_2$. However, evidence of antiferromagnetic (AFM) spin fluctuations are not obvious\cite{Yu_CM_11011017}.

These observations immediately lead to the question whether the superconducting properties of K$_{y}$Fe$_{2-x}$Se$_2$ is the same as that of the iron 
pnictides and $\alpha$-Fe$_{1-\delta}$Se, where the AFM spin fluctuations and interband interactions are believed important for the superconductivity. 
In K$_{0.8}$Fe$_{2-x}$Se$_2$, a large isotropic gap is suggested by the ARPES \cite{Feng_CM_10125980}, and preliminary NMR data give a bulk evidence 
of singlet pairing\cite{Yu_CM_11011017}. So far the superconducting properties are not intensively investigated in the ternary iron selenides, and 
more work is necessary in order to gain better understanding on the mechanism of the superconductivity.  

In this paper, we report our $^{77}$Se NMR Knight shift ($^{77}K_n$) and spin lattice relaxation rate ($1/^{77}T_1$) studies on high quality 
K$_{0.8}$Fe$_{2-x}$Se$_2$ single crystals ( $T_C$ ranging between 29.5-32 K). Below $T_C$, the Knight shift decreases with T. However, the 
$1/^{77}T_1$ does not show a coherence peak. In fact, the SLRR decreases rapidly down to $T_c/2$, which can be fit with a single isotropic gap $\Delta 
\approx (3.8\pm 0.5)k_B T_c$. These data gives bulk evidence for a strongly coupled superconductor with a singlet pairing and an isotropic gap. Below 
$T_c/2$, however, the $K_n$ ceases to decrease with T, and the $1/^{77}T_1$ shows a $1/^{77}T_1\sim T^2$ behavior. We discuss these low-temperature 
behaviors with possible intrinsic low energy excitations, or vortex core contributions.

\begin{figure}
\includegraphics[width=8cm, height=6cm]{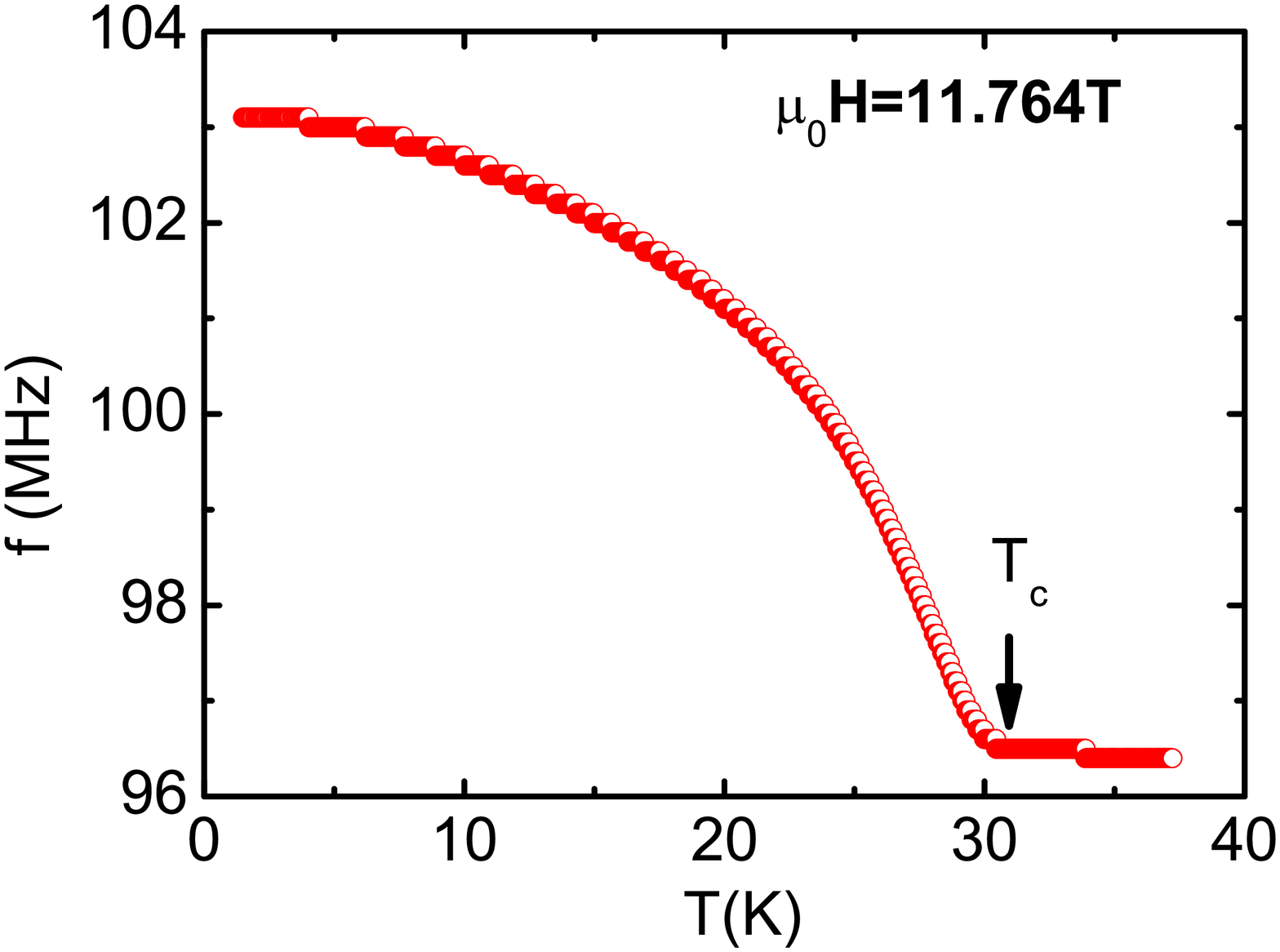}
\caption{\label{sus1}(color online) The resonance frequency $f$ of the NMR tank coil (with sample S2 inside) under a fixed tuning, which shows an 
onset $T_C$ of 32 K by the increase of $f$ due to the demagnetization effect. The frequency shift is given by $\Delta f(T) \propto -\chi_{ac}(T) $, 
where $\chi_{ac}(T)$ is the ac susceptibility of the NMR sample.}
\end{figure}
 
The K$_{0.8}$Fe$_{2-x}$Se$_2$ single crystals were synthesized by the flux-growth method \cite{Chen_CM_11010789}. For this NMR study, two large single 
crystals ($S1$ and $S2$) with a dimensions of 3$\times$2$\times$0.5mm$^3$ is chosen for the measurements. Sample $S1$ was also used for the normal 
state studies reported elsewhere \cite{Yu_CM_11011017}. Sample $S2$, from the same batch, is moderately grounded into small pieces to gain better RF 
penetration (see Fig.~\ref{sus1} for the susceptibility data). The $^{77}$Se Knight shift and the spin-lattice relaxation rate (SLRR) measurements 
were performed under a 12 Tesla magnetic field, and the superconducting transition is observed at 29.5 K for S1  \cite{Yu_CM_11011017} and 31 K for S2 
under field. We compared the data for sample S1 with $H//ab$ and $H//c$, and for sample S2 primarily sampling the spectra with $H//ab$. The Knight 
shift $K_n(T)$ is obtained from $K_n(T)=(f-f_0)/f_0$, where $f_0=\gamma B$ is the calculated frequency, and $f$ is the measured frequency of spectral 
maximum. The SLRR is deduced from the spin recovery after an inversion pulse, with the spectra collected with a CPMG sequence 
($(\frac{\pi}{2})_{x}-\pi _{y}-echo-\pi _{y}-echo...$) to maximize the signal-to-noise ratio. Our spin recovery data is well fit by the single 
exponential function $I(t)/I(0)=1-be^{-t/T_1}$. 

In general, the Knight shift is proportional to the electron density of states on the Fermi surface, which decrease below $T_c$. The spin-lattice 
relaxation rate $1/T_{1s}$ in the superconducting state are correlated with the normal state $1/T_{1n}$ by \cite{Hebel_PR_113_1504-1519, 
Hebel_PR_116_79}
\begin{eqnarray}
\frac{T^{-1}_{1s}}{T^{-1}_{1n}}=\frac{2}{k_BT}\displaystyle {\int^\infty_0 [\frac {N_S(E, \Delta)}{N_0(E)}]^2f(E)(1-f(E))\, dE }  &&\label{a3}
\end{eqnarray}
where $N_S(E, \Delta)$  and $N_0(E)$ are the electron density of states in the normal state and in the superconducting state respectively, and $f(E)$ 
is the Fermi-Dirac distribution function with $f(E)=1/(e^{-E/k_BT}+1)$.  

\begin{figure}
\includegraphics[width=8cm, height=6cm]{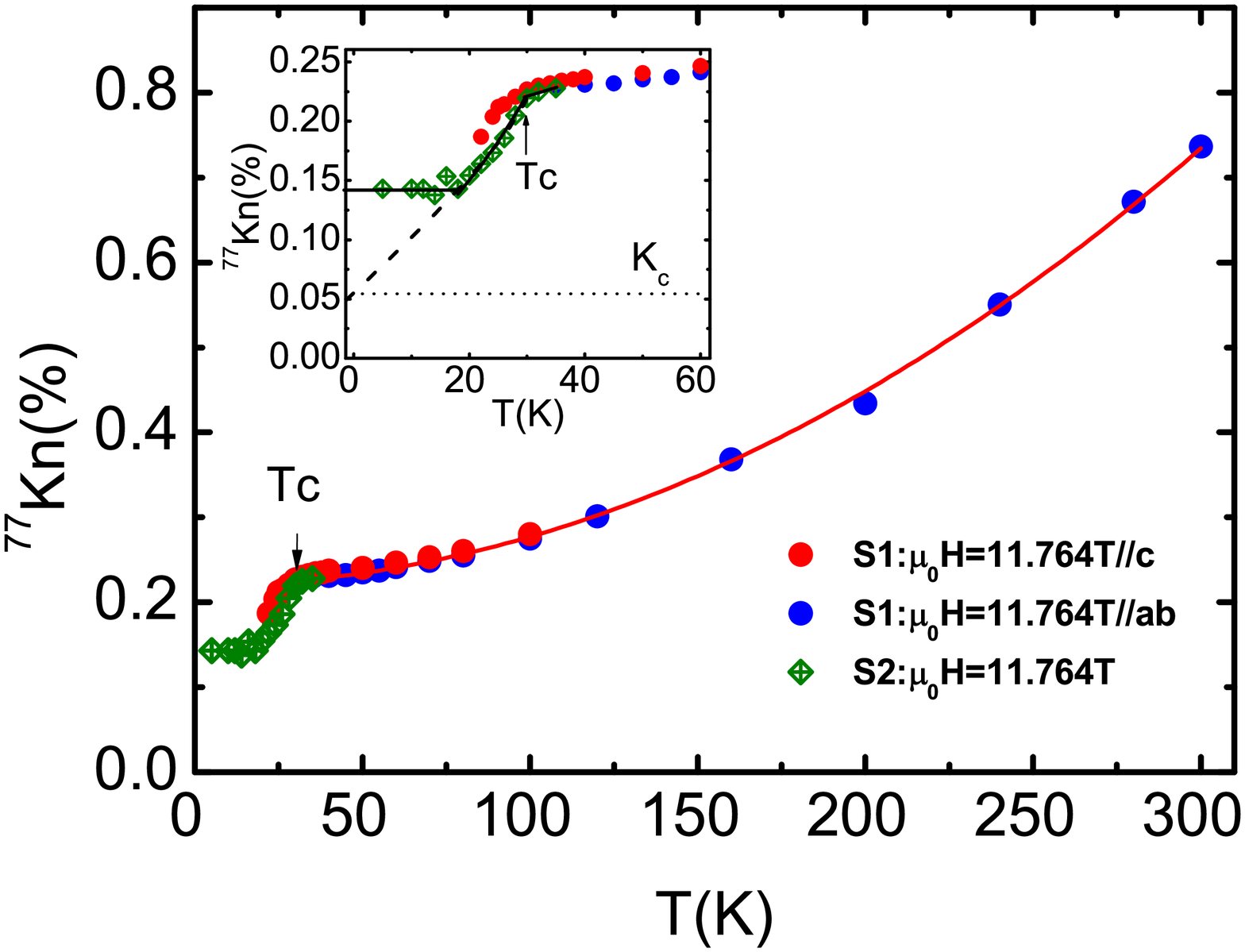}
\caption{\label{knight2}(color online) (a) The temperature dependence of the $^{77}$Se Knight shift $^{77}K_n$ for sample $S_1$ (with $H//ab$ and 
$H//c$) and for sample $S_2$. Inset: The enlarged view of the low-temperature Knight shift, which shows an onset $T_C$ about 30 K. The Chemical part 
of the Knight shift $K_c$, which does not change with temperature, is indicated by a dotted line.} 
\end{figure}

The temperature dependence of the Knight shift $^{77}K_n$ for two samples with different field orientations are shown in Fig.~\ref{knight2}. Above 
$T_c$, the Knight shift is isotropic, and decreases gradually as temperature drops \cite{Zhang_EPL}. Below $T_c$, the Knight shift drops faster with 
field along the $ab$-plane ($^{77}K^{ab}_n$) than with field along the $c$-axis ($^{77}K^{c}_n$) for sample $S1$, indicating an anisotropic $H_{C2}$. 
The $^{75}K^{ab}_n$ are consistent for sample S1 and S2, and we primarily discuss the results of sample $S2$ in the following.  In general, the Knight 
shift $K_n (T)$ is written as $K_n (T)=K_c+K_s(T)$, where $K_c$ is the chemical shift \cite{Yu_CM_11011017} and the $K_s(T)$ is the contribution from 
the spin part. Our analysis on $^{75}K_n (T)$ in the normal state gives a constant $^{77}K_c$ of $\sim 0.05\%$, and the $K_s(T)$ changing with 
temperature \cite{Yu_CM_11011017}.

A large decrease of $K_n$ is clearly seen from $T_c$ down to $T_c/2$, which suggests that the system is primarily a singlet superconductor. This is 
consistent with most iron arsenide superconductor. For comparison, the decrease of $K_n$ has not been reported by $\alpha$-FeSe system below $T_c$ to 
our knowledge \cite{Imai_prl_102_177005}.
  
\begin{figure}
\includegraphics[width=8cm, height=6cm]{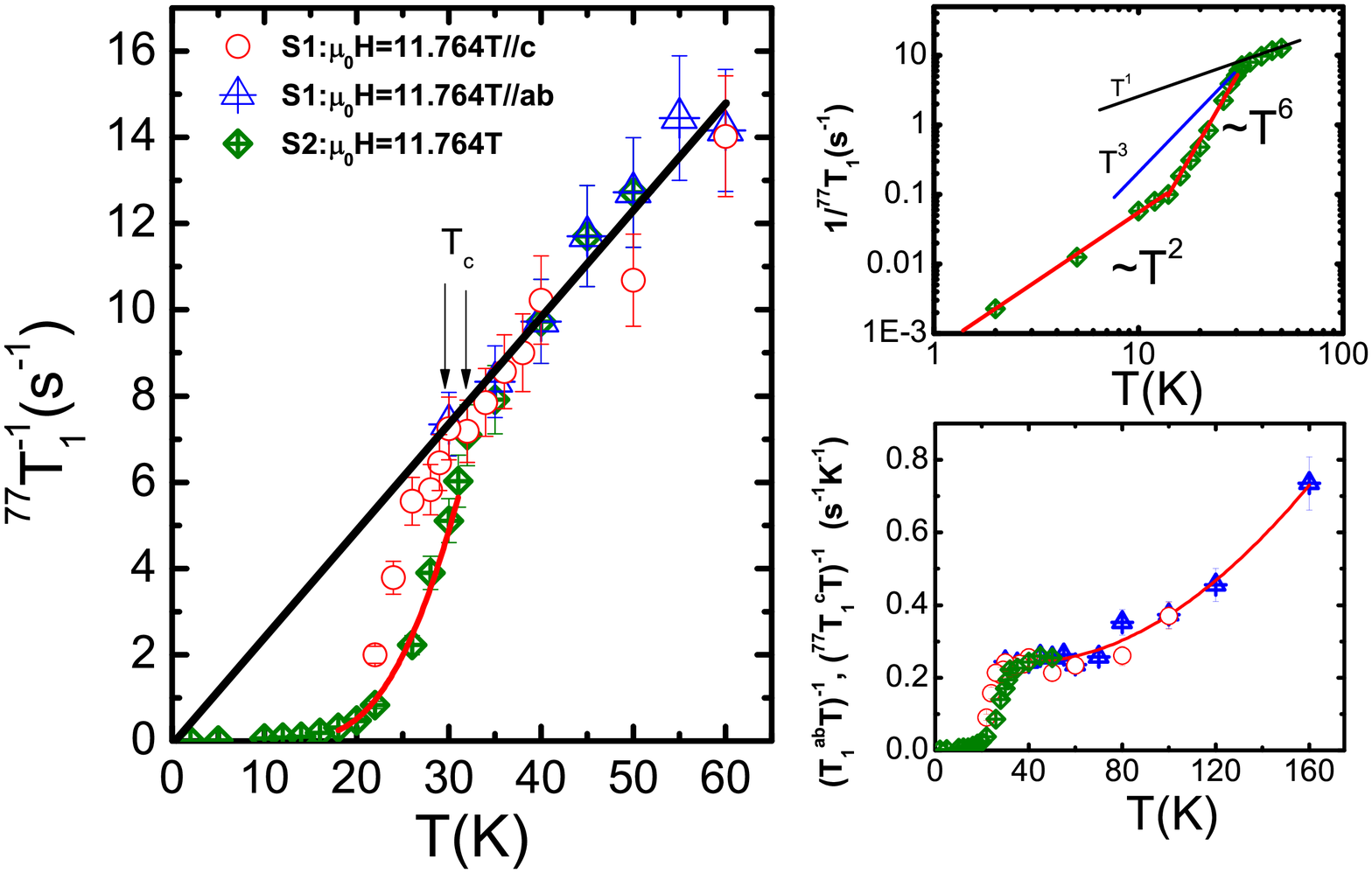}
\caption{\label{slrr3}(color online) (a) The temperature dependence of the spin-lattice rate  for sample S1 and for sample S2. The solid line is a 
fitting to the gap function $1/^{1}T^{ab}_1 = Ae^{-\Delta /k_BT}$. (b) The log-log plot of the spin-lattice rate $1/T^{ab}_1T$ for sample S2. (c) The 
$1/T_1T$ at difference temperatures, showing a Fermi-liquid-like behavior above $T_c$. }
\end{figure}

In order to gain more insight on the pairing symmetry, the low-energy excitations are studied by the spin-lattice relaxation rate, as shown in  
Fig.~\ref{slrr3}. The SLRR is isotropic above $T_c$ and anisotropic below $T_c$, and we only discuss $1/^{77}T^{ab}_1$ for sample S2 with $H//ab$. 
Above $T_c$, the SLRR is consistent with a canonical Fermi liquid with $^{77}T_1TK^2_s$ as a constant \cite{Yu_CM_11011017}. Below $T_c$, the 
$1/^{77}T^{ab}_1$ drops quickly with temperature. As shown in Fig.~\ref{slrr3}(b), two orders of magnitude difference in the SLRR is seen in a large 
temperature range from $T_c$ down to $T_c/2$. The temperature dependence is close to a power-law behavior with $1/^{77}T_1\sim T^6$, as shown in 
Fig.~\ref{slrr3}. Such a rapid decrease is much faster than the $1/T_1\sim T^3$, which suggests that the line node is absent or not obvious in 
K$_{0.8}$Fe$_{2-x}$Se$_2$.

For conventional superconductors, the superconducting gap $\Delta (T)$ increases from zero as temperature drops below $T_C$, which produces a 
Hebel-Slichter coherence peak from $T_c$ down to 0.8 $T_c$ \cite{Hebel_PR_113_1504-1519, Hebel_PR_116_79}. In K$_{0.8}$Fe$_{2-x}$Se$_2$, our data does 
not show evidence of a coherence peak. The absence of the coherence peak draws a similarity to other iron-based supercondutors, which will be 
discussed later. 

For an isotropic singlet superconductor, the SLRR far below Tc (Eq.~\ref{a3}) can be simplified $1/T_1\sim e^{-\Delta (T) k_BT}$, where $\Delta (T)$ 
is the superconducting gap. We fit the $1/^{77}T^{ab}_1$ with a a constant gap slightly below $T_c$, and $\Delta \approx 10.8\pm 2$ mev is obtained as 
shown by the fit in Fig.~\ref{slrr3}a. Such a large value produces $\Delta/k_BT_c \approx 3.8$, over two times of the BCS value $\Delta/k_BT_c= 1.75$. 
Our obtained value is very close to the ARPES data \cite{Feng_CM_11010462}, and gives the bulk evidence for a strongly coupled, isotropic 
superconductor. Here we assumed that the gap opens quickly with temperature to get the full value of $\Delta (0)\sim 4k_BT_c$ about 2 Kelvin below 
$T_c$, so that the coherence peak diminishes naturally. 

In all, the temperature behavior of the Knight shift and the SLRR data from $T_c$ down to $T_c/2$ suggest a strongly coupled, singlet 
superconductivity with an isotropic gap, which is similar to other iron pnictides \cite{Ding_EPL_83_47001, ZhouXJ_CPL}. From our data, we cannot give 
direct evidences for multiple superconducting gaps in K$_{0.8}$Fe$_2$Se$_2$. However, multiple gaps with close values in different bands should lead 
to a single gap behavior in the SLRR. Furthermore, the absence of the Hebel-Slichter coherence peak, well known in other iron pnictide 
superconductors, may be an indirect evidence for multiple gaps with strong impurity scattering effect \cite{Bang_PRB_79_054529}.

For iron pnictides, an $s_\pm$ gap symmetry has been suggested due to the interband interactions between the electron band and the hole band 
\cite{Mazin_PRL_101_057003, Kuroki_PRL_101_087004, Wang_PRL_102_047005, Cvetkovic_EPL_85_37002}. Antiferromagnetic spin fluctuations are probably also 
important for the superconductivity in both the iron pnictides \cite{Ning_PRL_104} and the $\alpha$-FeSe system\cite{Imai_prl_102_177005}. In 
K$_{0.8}$Fe$_{2-x}$Se$_2$, however, the hole band is not observed in the center of the Bruillouin zone due to heavy electron doping. The AFM 
fluctuations are not observed by NMR studies \cite{Yu_CM_11011017}. Then the issue rises how to connect the pairing symmetry with the band structure 
and the spin fluctuations in K$_{0.8}$Fe$_{2-x}$Se$_2$, since the superconducting properties are similar for different compounds. It is possible that  
other types of spin fluctuations are effective to the superconductivity \cite{Xiang_CM_10125536, Xiang_CM_10126015}.

Below $T_c/2$, The temperature behavior of the Knight shift and the SLRR are unexpected. In principle, $^{77}K_n(T)$ should drop to $K_c\approx 0.5\%$ 
as $T\rightarrow 0$ (or $K_s(T)\rightarrow 0$) for a singlet superconductor. However, the spin-part of the Knight shift $K^{ab}_s(T)$  below $T_c/2$ 
levels off with a residual value of $0.09\%$ (Fig.~\ref{knight2}), by taking off our estimated value of $K_c\approx 0.05\%$. Currently we are not 
aware whether the large residual $K_s$ as $T\rightarrow 0$ is intrinsic, or possibly due to vortex contributions.

The SLRR also shows a slower decrease with $1/T_1\sim T^2$ below $T_c/2$ (Fig.~\ref{slrr3}c). In fact, such behavior has also been reported in various 
iron-based superconductors. If we assume the low temperature behavior is intrinsic, low energy excitations are expected to raise the Knight shift and 
the SLRR rate. Such small value may indicate small gap minimum on the Fermi surface. Applying a two-gap function for the $1/^{77}T_1$, a second gap of 
$\Delta _2\sim 0.25 T_c$ with a very low electron density of states ($\sim 0.1\%$) is obtained. Alternatively, these behaviors could also be caused by 
vortex core contribution \cite{Hammerath_CM_09123681}, which is unlinked to the superconductivity regions. Measurements under different fields may be 
helpful to verify if the vortex contributions exists.    

To summarize, our NMR Knight shift study on K$_{0.8}$Fe$_{2-x}$Se$_2$ has shown the bulk evidence of singlet superconductivity. The spin-lattice 
relaxation data further support a strongly coupled superconductor with an isotropic gap $\Delta \approx 3.8 k_BT_c$. The coherence peak is not 
observed in the SLRR, and may be an indirect evidence for unconventional superconductivity. These observation draw a lot of similarities of the 
pairing symmetry among K$_{0.8}$Fe$_{2-x}$Se$_2$ and other iron-based superconductors, although their respective band structure and the magnetic 
correlations seem to be very different. We also observed a level off of the Knight shift and a low-power behavior of the SLRR below $T_c/2$, which are 
not understood.
  
The Authors acknowledge Professor Tao Li, Zhongyi Lu, Yaohua Su, Tao Xiang and Guangming Zhang for enlightening discussions. W.Y. and G.F.C. are 
supported by the National Natural Science Foundation of China (Grant Nos. 11074304 and 10974254) and the National Basic Research Program of China 
(Contract Nos. 2010CB923000 and 2011CBA00100).  


\end{document}